\documentclass[11pt]{article}
\usepackage{jheppub}
\usepackage{subfigure}

 \def\cO{{\cal O}}

\def\ben{\begin{equation}}
\def\een{\end{equation}}
\def\bea{\begin{eqnarray}}
\def\eea{\end{eqnarray}}
\def\vx{{\vec{x}}}

\def\nn{\nonumber}
\def\cD{{\cal{D}}}
\def\cG{{\cal{G}}}
\def\cA{{\cal{A}}}
\def\cB{{\cal{B}}}
\def\br{{\bar{r}}}
\def\brho{{\bar{\rho}}}
\def\bG{{\bar{G}}}
\def\bV{{\bar{V}}}
\def\cP{{\cal{P}}}
\def\tchi{{\bar{\chi}}}
\def\bxi{{\bar{\xi}}}
\def\bB{{\bar{B}}}
\def\bC{{\bar{C}}}
\def\blambda{{\bar{\lambda}}}
\def\bc{{\bar{c}}}
\def\txi{{\tilde{\xi}}}
\def\cQ{{\cal{Q}}}

\begin{document}
\title{Quantum Quench and Double Trace Couplings}

\author{Pallab Basu $^{(a)}$,} 
\emailAdd{pallabbasu@gmail.com} 
\affiliation{$^{(a)}$ International Center for Theoretical Sciences,
Tata Institute of Fundamental Research,
Bangalore 560012, INDIA}

\author{Diptarka Das$^{(b)}$,}
\emailAdd{diptarka.das@uky.edu}
\affiliation{$^{(b)}$Department of Physics and Astronomy, University of Kentucky, Lexington, KY 40506, USA}

\author{Sumit R. Das $^{(b)}$}
\emailAdd{das@pa.uky.edu}

\author{and Krishnendu Sengupta $^{(c)}$}
\emailAdd{ksengupta1@gmail.com}
\affiliation{$^{(c)}$Department of Theoretical Physics, Indian Association for the Cultivation of Science, Jadavpur, Kolkata 700032, INDIA}

%
\abstract{We consider quantum quench by a time dependent double trace coupling
in a strongly coupled large N field theory which has a gravity dual
via the AdS/CFT correspondence. The bulk theory contains a self
coupled neutral scalar field coupled to gravity with negative
cosmological constant. We study the scalar dynamics in the probe
approximation in two backgrounds: AdS soliton and AdS black brane.
In either case we find that in equilibrium there is a critical phase
transition at a {\em negative} value of the double trace coupling
$\kappa$ below which the scalar condenses. For a slowly varying
homogeneous time dependent coupling crossing the critical point, we
show that the dynamics in the critical region is dominated by a
single mode of the bulk field. This mode satisfies a Landau-Ginsburg
equation with a time dependent mass, and leads to Kibble Zurek type
scaling behavior. For the AdS soliton the system is non-dissipative
and has $z=1$, while for the black brane one has dissipative $z=2$
dynamics. We also discuss the features of a holographic model 
which would describe the non-equilibrium dynamics around quantum critical points with
arbitrary dynamical critical exponent $z$ and correlation length
exponent $\nu$. These analytical results are supported by direct
numerical solutions.}

\maketitle
\section{Introduction and Summary}

There has been a lot of interest in understanding the problem of
thermal or quantum quench \cite{sengupta, CC} using gauge-gravity
duality \cite{Maldacena}. One set of works concentrate on the question
of thermalization by horizon formation
\cite{thermalization1}-\cite{otherapparent} and possible resolutions
of spacelike singularities \cite{holocosmo}. Recently there have been
several studies of holographic quench which involve critical
points. In \cite{basu-das} two of us initiated the study of
holographic quench across finite temperature and finite chemical
potential critical points, and found hints of a mechanism which gives
rise to Kibble-Zurek scaling in critical dynamics
\cite{kibblezurek,sengupta}. This mechanism was confirmed
for a zero temperature but nonzero chemical potential quantum critical
point in \cite{bddn}. In slightly different directions \cite{murata, bhaseen}
studied relaxation dynamics following a thermal quench from a broken
symmetry phase and \cite{buchel} studied scaling behavior of final
values of observables due to a thermal quench. Quantum quench in
solvable large-N field theories without the use of gauge-gravity
duality has been studied in
\cite{sotiriadis,dsengupta,gubsersondhi,mandalmorita}.

In \cite{basu-das} and \cite{bddn} the quench was due to a homogeneous
time dependent source for a scalar order parameter which translates to
a time dependent Dirichlet boundary condition on the strongly
self-coupled bulk scalar field. The other parameters in the theory
were tuned such that in the absence of a source the theory is
critical. The dynamics was then studied in the probe approximation by
considering a source which is slowly varying at early and late times
and which crosses zero (i.e. the location of the critical point) at
some intermediate time. In this setup scaling behavior appears due to
a few key facts

\begin{itemize}

\item{} At the equilibrium critical point the linearized bulk equation
  of motion for the scalar has a zero mode. This results in a
  breakdown of adiabaticity when the source becomes small
  characterized by a power law in the rate of change of the source
  $v$.

\item{} In the critical region, and only in this region, there a new
  expansion for small $v$. This is an expansion in fractional powers
  of $v$, with exponents determined by the equilibrium critical
  exponents.

\item{} In the lowest order of this expansion in fractional powers of
  $v$, the bulk dynamics is dominated by the zero mode. This zero mode
  then satisfies an ordinary differential equation which is basically
  the dynamics of the order parameter. In this equation, the boundary
  condition appears as a source term. This equation has a scaling
  solution displaying Kibble-Zurek scaling.

\end{itemize}

The setup in \cite{basu-das} and \cite{bddn} involved a nonzero
chemical potential and/or nonzero temperature. The background in
\cite{basu-das} is a charged black brane with a neutral self coupled
scalar \cite{liu1}, while that in \cite{bddn} is an AdS soliton with a
constant gauge field and a self coupled scalar - a variation of the
setup of \cite{nishioka}. In both cases the resulting dynamics of the
order parameter is non-relativistic with dynamical critical exponent
$z=2$, even though the underlying bulk dynamics is relativistic.  It
is possible that the zero temperature limit of the setup of
\cite{basu-das} may lead to a $z=1$ dynamics. However the zero
temperature limit the phase transition found in \cite{liu1} and probed
in \cite{basu-das} becomes a Berezinski-Kosterlitz-Thouless transition
and we were not able to get any analytic handle on the dynamics.

So far all studies of quantum or thermal quench using holographic
methods have dealt with time dependent external sources. A useful
example to keep in mind is a magnet in the presence of a time
dependent magnetic field. Critical dynamics can be then studied by
tuning the temperature to the critical value. In a Landau-Ginsburg
language this corresponds to a time dependent inhomogeneous term in
the LG equation. In many situations, this is not a natural thing to
do. For example in a superconductor an external source for the order
parameter is not very natural, though it can be achieved by
considering junctions. On the other hand, the standard tuning
parameter in a critical transition is the term in a LG hamiltonian
which is quadratic in the order parameter: we will call this a LG
mass term.

In this paper we initiate the study of quench by such a time dependent
LG mass using holographic techniques.  While studying holographic
quench with time dependent external source is straightforward because
it maps to a time dependent boundary condition for the dual field, a
time dependent LG mass quench would involve addition of a {\em double
  trace deformation} with a time dependent coefficient,
$\kappa(t)$. As is well known this implies a modified boundary
condition for the bulk scalar \cite{doubletrace}.  In equilibrium
\cite{fhr} found that for a class of scalar potentials, there is a
critical phase transition at $\kappa = \kappa_c$ where $\kappa_c < 0$.
Naively, from the field theory viewpoint, a deformation with negative
$\kappa$ appears to lead to an instability. However it has been shown
in \cite{fhr} and \cite{fhr2} this is not necessarily correct -
typically there is a stable ground state with scalar hair for $\kappa
< \kappa_c$. For vanishing temperature and vanishing chemical
potential $\kappa_c = 0$, while for a nonzero temperature (i.e a black
hole background) $T$ one has
$\kappa_c \propto T$. In the following we will show, not surprisingly, that there is a
similar transition when the background is a $AdS$ soliton.

We consider the simplest situation where such a transition occurs. The
bulk action is given in $L_{AdS}=1$ units
\ben S= \int
d^{d+2}x~\sqrt{g} \left[ \frac{1}{8\pi G_N} \left(R + d(d+1) \right) -
  \frac{1}{\lambda} \left( (\nabla \phi)^2 +m^2 \phi^2
  + V(\phi) \right) \right]
\label{1-1}
\een
where $\phi$ is a neutral bulk scalar. We will consider the limit
$\lambda \gg G_N$ so that the scalar can be treated as a probe field
whose dynamics does not affect the gravity background. We will
consider potentials $V(\phi)$ which have a power series expansion in
$\phi$. As will become clear soon, the critical behavior is determined
by the leading nonlinearity in $V(\phi)$, so it would be sufficient to
consider monomials.

First we study the equilibrium transition in three such
backgrounds. The first is pure $AdS_{d+2}$, which is the relevant geometry
when all the spatial directions are noncompact,
\ben ds^2 = r^2(-dt^2
+ d\vx^2 +dw^2)+ \frac{dr^2}{r^2}
\label{1-2}
\een
The second is a $AdS_{d+2}$ soliton which is the relevant
geometry when one of the spatial directions, $w$ is compact with some
radius $R_0$,
\bea
ds^2 & = & r^2(-dt^2 + d\vx^2 + f_s(r) du^2)+ \frac{dr^2}{r^2 f_s(r)}
\nn \\
f_s(r) & = & 1 - \left( \frac{r_0}{r} \right)^{d+1},~~~~~~r_0 = \frac{4\pi}{(d+1)R_0}
\label{1-3}
\eea
The third is a $AdS_{d+2}$ black brane with all boundary directions
non-compact. The metric is
\bea
ds^2 & = & - r^2 f_b(r) dt^2 + r^2(d\vx^2 + dw^2)+ \frac{dr^2}{r^2 f_b(r)}\nn \\
f_b(r) & = & 1 - \left( \frac{\br_0}{r} \right)^{d+1},~~~~~~\br_0 =
\frac{4\pi T}{(d+1)}
\label{1-3a}
\eea

In all these cases the asymptotic form of the solution for
the scalar has the form
\ben
\phi(r,\vx,t,u) \rightarrow r^{-\Delta_-}\cA(\vx,u,t) \left( 1
+ O(1/r^2) \right) + r^{-\Delta_+}\cB(\vx,u,t) \left( 1
+ O(1/r^2) \right)
\label{1-4}
\een
{\em provided} the solution becomes small near the boundary. In (\ref{1-4})
\ben
\Delta_\pm = (d+1)/2 \pm \sqrt{(d+1)^2/4 + m^2}
\label{1-5}
\een
We will work in the mass range $ -(d+1)^2/4  \leq m^2 \leq
-(d+1)^2/4 +1 $ so that we have two possible quantizations \cite{alternative}:
the standard quantization with $\cA$ as the source and the
alternative quantization with $\cB$ as the source. The dimensions of
the dual operator $\cO$ in these two quantizations are $\Delta_+$ and
$\Delta_-$ respectively.

It is also possible to impose boundary conditions
\ben
\cB(\vx,u,t) = \kappa(\vx,u,t) \cA(\vx,u,t)
\label{1-6}
\een
As is well known, this corresponds to addition of a term \cite{doubletrace}
\ben
\int d^{d+1}x~\kappa \cO^2
\label{1-7}
\een
to the field theory action.

We will first show explicitly for suitable potentials that for a
constant $\kappa$ all these backgrounds admit critical points. For
$AdS_{d+2}$ the critical value is $\kappa = 0$ : for $\kappa < 0$
there is a nontrivial solution of the equations of motion which is
regular everywhere and satisfies the specified boundary conditions.
This means that for the dual operator $\langle\cO\rangle \neq 0$.
Near the critical point we verify that \ben \langle\cO\rangle \sim
(-\kappa)^{\Delta_-/(\Delta_+-\Delta_-)} \label{1-8} \een as
required by scale symmetry. The scaling behavior is independent of
the nature of the potential whose properties enter only in the
overall coefficient.

For the $AdS_{d+2}$ soliton as well as the $AdS_{d+2}$ black brane,
the critical value of $\kappa$ is at
some finite value $\kappa_c <0$ and the condensate appears for
$\kappa < \kappa_c$. This is shown by a direct numerical solution.
It turns out that the value of $\kappa_c$ can be determined
analytically in closed form, following the treatment of \cite{fhr} and
is independent of the nature of the non-linearity.
As is usual in such situations, there is a zero mode of the
linearized equation at $\kappa = \kappa_c$ : here the zero mode has
a closed form in terms of hypergeometric functions.  We verify that
our numerical solution for $\phi^4$ and $\phi^6$ potentials agrees
with this. The critical exponent can be also determined
analytically. When the leading nonlinearity is $\phi^{n+1}$ , one
gets \ben \langle\cO\rangle \sim (\kappa_c - \kappa)^{1/(n-1)}
\label{1-9} \een Note that in standard notation the critical
exponent $\beta$ is given by \ben \beta = 1/(n-1) \een Our numerical
results are consistent with the behavior (\ref{1-9}). We also verify
numerically that the critical behavior is indeed determined by the
leading non-linearity.

We then consider the response of the system to a time dependent but
homogeneous $\kappa (t)$ for the $AdS_{d+2}$ soliton and $AdS_{d+2}$
black brane backgrounds, staying in the probe approximation. For these
backgrounds, the radius of the compact dimension (for the soliton) or
the temperature (for the black brane) provides a scale, so that we can
meaningfully talk about slow and fast quenches. We concentrate on slow
quench starting deep in the ordered phase, crossing the critical point
$\kappa_c$ at some intermediate point and asymptoting to some other
constant value at late times. Following the lines of
\cite{basu-das,bddn} we study the breakdown of adiabaticity and show
that in a way similar to these works the critical region is
characterized by an expansion in fractional powers of the rate and by
the dominance of the zero mode. For a fast quench we expect
a chaotic behavior to set in \cite{Basu:2013vva}. Unlike these previous works, the
function $\kappa (t)$ now appears as a time dependent mass term in the
effective LG dynamics of the zero mode and hence the order
parameter. This is consistent with the fact that in the field theory,
$\kappa(t)$ is indeed the coefficient of $\cO^2$.

The ensuing critical dynamics for the soliton and the black brane are
different. For the soliton, the dynamics is relativistic and non-dissipative. This is
expected since in the field theory is at zero temperature and there is
no chemical potential. The dynamics in the black brane background has
$z=2$ and is dissipative, as would be expected for a finite
temperature situation.

Finally we solve the time evolution numerically and provide evidence
for the scaling behavior discussed above.

In Section 2 we set up the equilibrium problems, show the existence of
the critical point for negative constant $\kappa$ and derive the
critical exponents. Sections 3 and 4 deal with quantum quench due to a
time dependent $\kappa (t)$ for the soliton and black brane backgrounds respectively. In section 5 we present our numerical results. In section 6 we discuss arbitrary critical exponents $z$ and $\nu$ and the relationship of our scaling solutions with standard Kibble-Zurek scaling. Section 7 contains brief remarks and the appendix discusses the solution of a toy model which justifies some key ingredients in our discussion of section 4.

\section{The equilibrium critical point}

In the probe approximation the only relevant equation we need to solve
is the scalar field equation. For the backgrounds (\ref{1-2}) or
(\ref{1-3}) and fields which depend only on $t$ and $r$ this equation
is
\ben -\frac{1}{h(r)}
\partial_t^2 \phi + \frac{1}{r^{d-2}}\partial_r
(r^{d}g(r)\partial_r \phi) -m^2r^2 \phi - r^2 V^\prime(\phi) = 0
\label{2-1}
\een
where
\ben
g(r) = \begin{cases} r^2, & \mbox{for } AdS_{d+2} \\
 r^2 f_s(r) & \mbox{for } AdS_{d+2} \mbox{ soliton} \\
 r^2 f_b(r) & \mbox{for } AdS_{d+2} \mbox{black brane}
\end{cases}
\label{2-2}
\een
and
\ben
h(r) = \begin{cases} 1 & \mbox{for }AdS_{d+2} \mbox{and }AdS_{d+2} \mbox{ soliton}\\
 f_b(r) & \mbox{for }AdS_{d+2} \mbox{ black brane}
\end{cases}
\label{2-2a}
\een
We first need to find static solutions of (\ref{2-1}) which are regular in the interior and which satisfy the boundary condition (\ref{1-6}) at the boundary (with constant $\cA,\cB,\kappa$).

\subsection{Pure $AdS_{d+2}$}

For pure $AdS_{d+2}$ and a $\phi^4$ potential regularity
means that the value of the field at $r=0$ is fixed to the attractor value
\ben
\phi(r=0)_{AdS} = \sqrt{-m^2}
\label{2-3}
\een
To find a solution to the non-linear equation consider integrating the equation by imposing the condition at small $\epsilon$
\ben
\phi(\epsilon) = \phi (r=0)_{AdS} - c \epsilon^{\Delta_v}~~c > 0~~~\Delta_v = \sqrt{(d+1)^2/4 - 2m^2}-(d+1)/4
\label{2-4}
\een
This form is dictated by the solution near $r=0$ where the departure
from the attractor value is small so that the equation can be
linearized. The solution to the full nonlinear equation may be
therefore written as $\phi(r,c)$, which gives us a one parameter class
of solutions. However the equation has a scaling symmetry under
\ben
r \rightarrow \lambda r~~~~~~~~~~~~~~\phi \rightarrow \phi
\label{2-5}
\een
which immediately implies that the solution satisfies
\ben
\phi_{AdS}(r,c) = \phi_{AdS} (rc^{-1/\Delta_v},1)
\label{2-6}
\een
The solution near the boundary $r = \infty$ is of the form
(\ref{1-4}) with constant $\cA$ and $\cB$ - the scaling symmetry then implies
\ben
|\cA |  \sim c^{\frac{\Delta_-}{\Delta_v}}~~~~
| \cB  | \sim c^{\frac{\Delta_+}{\Delta_v}}~~\implies~~
| \kappa | \sim c^{\frac{\Delta_+-\Delta_-}{\Delta_v}}
\label{2-7}
\een
In alternative quantization, $\cA$ is the expectation value of the
dual operator and the above relations immediately implies (\ref{1-8}).

The solution $\phi(r,c)$ can be found easily by numerically solving
the nonlinear equation. We find that there is a nonsingular solution
for any negative $\kappa$ which satisfies the above scaling behavior.

Note that the scaling argument given above does not depend on the
potential being $\phi^4$, and is valid for an arbitrary potential
$V(\phi)$. The value of the attractor is generally given by
\ben
m^2 \phi + V^\prime(\phi) = 0
\een
and the behavior for small $r$ becomes a bit complicated, though still
determined by a linear equation. Nevertheless the same scaling
behavior (\ref{1-8}) would follow. The numerical coefficient will of
course depend on the details of the potential.

\subsection{$AdS_{d+2}$ soliton}

For the $AdS_{d+2}$ soliton (\ref{1-3}) regularity at the tip
$r=r_0=1$ implies that the field can attain any value $\phi_0$ at
$r=r_0$ while the derivative is given by
\ben
\frac{d\phi}{dr}(r=r_0) = \frac{1}{d+1}\left[ m^2 \phi_0 +
  V^\prime(\phi_0) \right]
\label{2-8}
\een
The static solution may be now obtained by starting at some $\phi_0$
and integrating out to $r=\infty$. As we will see below, straightforward numerical
integration then shows that a non-trivial regular solution exists only
when $\kappa < \kappa_c$ where the critical value $\kappa_c$ is a
negative number to be determined shortly.

In the rest of the paper we will use $r_0=1$ units

As is usual in such cases (e.g. for holographic superconductors
\cite{hhh} ) the trivial solution with $\phi = 0$ is in fact unstable
for $\kappa < \kappa_c$. To see this, let us write the linearized
equation of motion as
\bea
-\partial_t^2 \phi & = &{\tilde {\cD}} \phi \nn \\ {\tilde{\cD}}
& \equiv &- r^{2-d}\frac{\partial}{\partial
  r} \left( r^{d+2}f_s(r) \frac{\partial}{\partial r}\right) +m^2 r^2
\label{2-9}
\eea
This equation can be cast into a Schrodinger form by changing
coordinates to $\rho$ \ben \rho(r) = \int_r^\infty
\frac{ds}{s^2 f_s^{1/2}(s)}
\label{2-10}
\een
and redefining the field to $\psi(\rho,t)$
\ben
\phi(r,t)
= \frac{1}{[r(\rho)]^{\frac{d}{2}-1}}\left(\frac{d\rho}{dr}
\right)^{1/2} \psi(\rho,t)
\label{2-12}
\een
Note that
\bea
\rho & \sim & 1/r~~~~~~~~~~~~~r \rightarrow \infty \nn \\
\rho & \sim & \rho_\star + \frac{2\sqrt{r-1}}{\sqrt{d+1}}~~~~~~~r \rightarrow 1
\label{2-11}
\eea where $\rho_\star$ is finite. For $d=3$ we get $\rho_\star =
1.311$.
Using the
explicit form of $f_s(r)$ the equation (\ref{2-9}) becomes
\ben
-\partial_t^2 \psi = \cD \psi
\label{2-11a}
\een
where
\bea \cD & = & -\partial_\rho^2 + V_0(\rho)
\nn \\ V_0(\rho) & = & m^2 r^2+ \frac{4d
  [(d+2)r^{2d+2}-(d+3)r^{d+1}]-(d-1)^2}{16r^{d-1}(r^{d+1}-1)}
\label{2-13}
\eea
This operator appeared in \cite{bddn} where it was shown that
with boundary conditions corresponding to either standard or
alternative quantization this has a positive spectrum. However with
the modified boundary condition $\cB = \kappa \cA$ with $\kappa <
0$ this is no longer true. In fact there is a specific value of
$\kappa$ where the operator $\cD$ has a zero mode. The equation ${\tilde{\cD}}
\phi = 0$ (which is equivalent to $\cD \psi = 0$)
is in fact the same as equation (B.1) in \cite{fhr} and we
can borrow the results. The solution which is regular at $r=1$ is
given by \ben \phi_0 (r) = A \left( r^{-\Delta_-} F^2_1
\left[\frac{\Delta_-}{d+1}, \frac{\Delta_-}{d+1},
  \frac{2\Delta_-}{d+1},r^{-(d+1)}\right] + B r^{-\Delta_+} F^2_1
\left[\frac{\Delta_+}{d+1}, \frac{\Delta_+}{d+1},
  \frac{2\Delta_+}{d+1},r^{-(d+1)}\right] \right)
\label{2-14a}
\een
where
\ben
B = -\frac{\Gamma(\frac{2\Delta_-}{d+1})[\Gamma(1-\frac{\Delta_-}{d+1}]^2}{\Gamma( 2 - \frac{2\Delta_-}{d+1})[\Gamma(\frac{\Delta_-}{d+1})]^2}.
\label{2-14b}
\een
The asymptotic expansion of this solution at $r=\infty$ can be read
off trivially. Clearly the $\kappa$ for this solution is $\kappa =
B$. This must be the critical value, $\kappa_c$ which is thus
determined to be \ben \kappa_c = - B~( r_0)^{d+1-2\Delta_-}
\label{2-14}
\een
where we have restored factors of $r_0$. The zero mode $\phi_0$ will
play a key role in what follows.

For $\kappa < \kappa_c$ the operator has negative eigenvalues which
implies that the trivial solution is unstable.

In the following we will also need the behavior of the lowest
eigenvalue $\lambda_0$ of $\cD$ for $\kappa < \kappa_c$. Generically
one would expect that this would vanish linearly, 
 \ben \lambda_0(\kappa) = - c_0 (\kappa_c
- \kappa).
\label{2-20}
\een
We have checked this numerically for $d=3$ and $m^2 = -15/4$ and obtained $c_0 = 0.762589$.  We also checked that $\kappa_c = -0.495$ which is consistent with (\ref{2-14}) and {\ref{2-14b}). This behavior will be important in the dynamics.

\subsubsection{Effect of non-linearity}

We now consider the effect of non-linearity in the static solution. Consider a $Z_2$ invariant potential of the form \footnote{See \cite{Kiritsis:2012ma} for a discussion of scalar effective potential in $AdS$ background.}
\begin{equation}
V(\phi)=\sum_{q=2}^{\infty} \lambda_q \phi^{2q}.
\end{equation}
For simplicity we will assume all the $\lambda_q$'s to be positive. 
We want to find solutions of the full nonlinear equation with specified boundary conditions at $r = \infty$ and regular in the interior.
Such solutions can be constructed by numerical
integration starting with a given value of $\phi_0$ and obtaining
the solution $\phi (r ; \phi_0)$ from which the leading and
subleading terms in the asymptotic expansion, $\cA$ and $\cB$ can be
calculated, thus determining $\kappa (\phi_0)$.
In all the cases we have studied, the solution is trivial for any $\kappa > \kappa_c $ while for $\kappa < \kappa_c$ there is a nontrivial solution,
leading to a nonzero order parameter in the boundary theory.
It is expected (and we can also verify numerically) that $\kappa_c$ is only a function of $m^2$ and it is independent of $\lambda_q$.
Furthermore, as is usual in mean field theory, the critical exponent is determined by the leading nonlinearity. For example, if lowest nonvanishing term is $V(\phi)=\frac{1}{4} \phi^4$ then one expects
 \ben 
 \langle\cO\rangle_{soliton} \sim (\kappa_c - \kappa)^{1/2}.
\label{2-15} 
\een 
This is standard mean field behavior. The exponent should not be affected by the presence of nonvanishing $\lambda_q$ with $q > 2$.

Figure (\ref{order-parameter-phi4phi6}) shows the result of a numerical
solution of the static equations of motion for $ d=3, m^2 = -15/4$ for two potenitals : (i) $\lambda_2 =1$ with all the other $\lambda_q$ vanishing and (ii) $\lambda_2 = 1, \lambda_3 = 20$ with the other $\lambda_q$ vanishing
 
\begin{figure}[htbp]
    \centering \includegraphics[width=10cm]{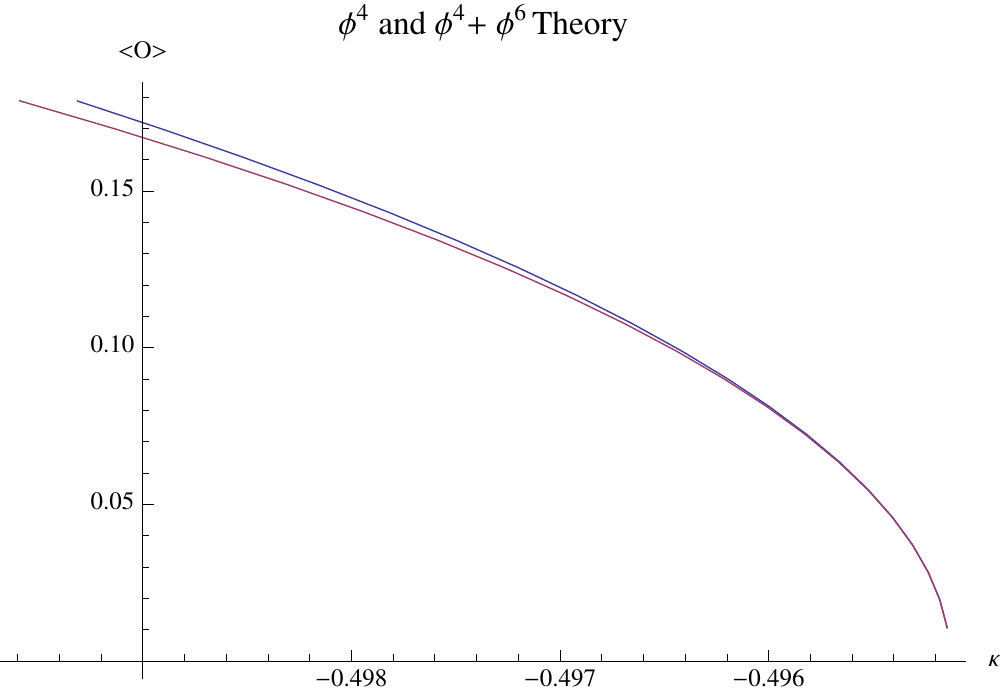}
    \caption{The order parameter as a function of $\kappa$ for
          $\phi^4$ (blue) and $\phi^4 +\phi^6$ (red) theory. The critical value is around -0.495129 }
    \label{order-parameter-phi4phi6}
\end{figure}

The critical coupling $\kappa_c$ is found to be $\kappa_c = -0.495129$ which is the same for both potentials and consistent with (\ref{2-14b}) and (\ref{2-14}) for this value of $d, m^2$. Clearly the behavior of the order parameter near the critical point is the same for both potentials while the behavior differs far away from the critical point. Figure (\ref{order-parameter-scaling-phi4phi6}) shows the determination of the critical exponent for bth potentials.

If the leading non-vanishing non-linear term is of $O(\phi^{n+1})$, i.e. $V(\phi)=\frac{1}{n+1} \phi^{n+1}+\cdots$, then we get,
\ben
\langle\cO\rangle_{soliton} \sim (\kappa_c - \kappa)^{\beta} 
\label{2-16d}
\een
where $\beta = \frac{1}{(n-1)}$. This is standard mean field multicritical behavior (for a numerical verification see Fig \ref{order-parameter-scaling-phi6.pdf}).

\begin{figure}[htbp]
    \centering
    \includegraphics[width=9cm]{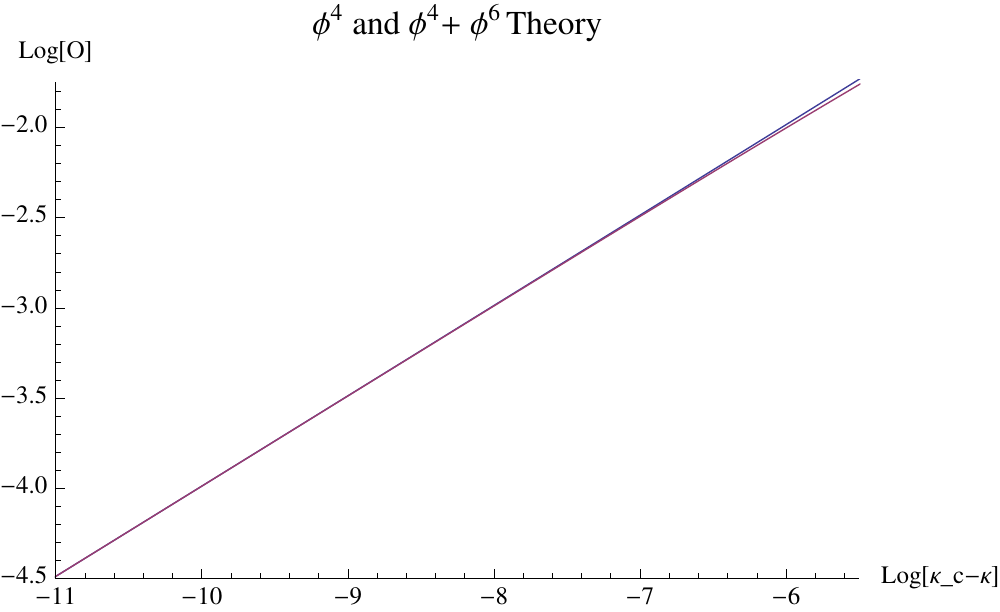}
    \caption{Plot of $\log \langle\cO\rangle$ versus $\log (\kappa_c - \kappa)$ for $\phi^4$ (blue) and $\phi^4 +\phi^6$ (red) theory. The fit for blue line is, $\log\langle \cO \rangle = 1.01802+ 0.500572 \log(\kappa_c - \kappa)$ and that for the red line is $\log\langle \cO \rangle = 0.96698 + 0.495077 \log(\kappa_c - \kappa)$}
    \label{order-parameter-scaling-phi4phi6}
\end{figure}

The critical exponent in fact follows from the equation itself. In
terms of the redefined field $\psi(\rho,t)$ ( see Eqs. (\ref{2-9}) -
(\ref{2-13})), the static equation of motion is 
\ben
\cD \psi + G(\rho) \psi^n = 0
\label{2-16a}
\een where 
\ben G(\rho) \equiv
\frac{r^{2-d}}{f_s(r)^{1/2}} \label{2-17} \een 
Near $\kappa =\kappa_c$ the solution itself is small and may be expanded as \ben
\psi(\rho;\kappa) = \epsilon^\beta \left( \psi^{(0)}(\rho) + \epsilon
\psi^{(1)}(\rho) + \epsilon^2 \psi^{(2)}(\rho) + \cdots \right) \label{2-18}
\een where \ben \epsilon \equiv (\kappa_c - \kappa) \label{2-19}
\een where the number $\beta$ has to be determined by substituting
the expansion in (\ref{2-17}) and equating terms order by order in
$\epsilon$. This may be easily seen to determine $\beta =  \frac{1}{(n-1)}$.

\begin{figure}[htbp]
\begin{minipage}[b]{0.49\linewidth}
\centering
\includegraphics[width=\textwidth]{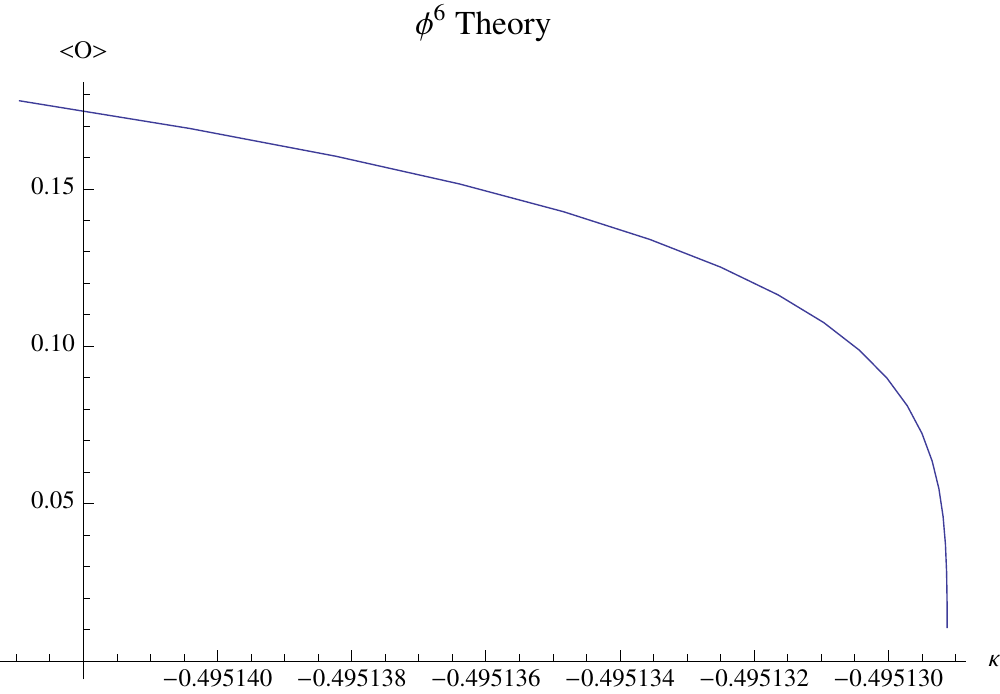}
\caption{The order parameter as a function of $\kappa$ for
          $\phi^6$ theory. The critical value is around -0.495129.}
\label{order-parameter-phi6}
\end{minipage}
\hspace{0.5cm}
\begin{minipage}[b]{0.49\linewidth}
\centering
\includegraphics[width=\textwidth]{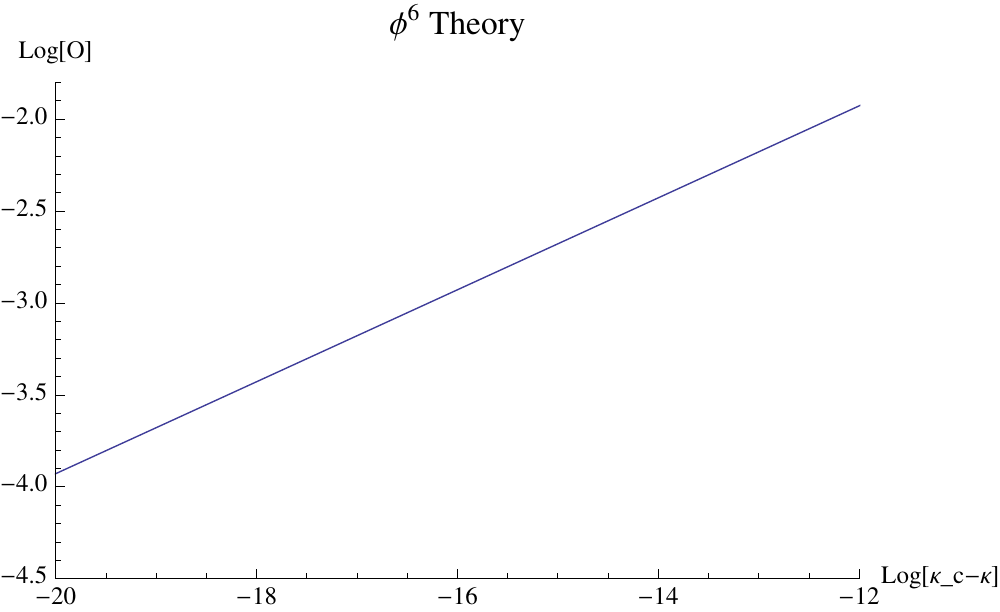}
\caption{Plot of $\log \langle\cO\rangle$ versus $\log (\kappa_c -
          \kappa)$ for $\phi^6$ theory. The fit is $\log\langle \cO \rangle = 1.07161+ 0.250041 \log(\kappa_c - \kappa)$ }
\label{order-parameter-scaling-phi6.pdf}
\end{minipage}
\end{figure}

\subsection{$AdS_{d+2}$ Black Brane}

The equilibrium solutions for the $AdS_{d+2}$ black brane are
identical to those for the $AdS_{d+2}$ soliton with the replacement
$r_0 \rightarrow \br_0$. This is clear from the
full equation (\ref{2-1}) and the form of the functions $f_b(r)$ and
$f_s(r)$ in (\ref{1-3}) and (\ref{1-3a}).

However, the passage to the Schrodinger form of the equations is
different, which leads to a different dynamics. As explained below, it
is useful to use Eddington-Finkelstein coordinates which are regular
at the horizon,
\ben
u= \brho - t~~~~~~~d\brho =  -\frac{dr}{r^2 f_b(r)}
\label{2-31}
\een
so that the metric becomes
\ben
ds^2 = r^2 (d\vx^2+dw^2) - 2 du dr - r^2 f_b(r) du^2
\label{2-32}
\een
In terms of fields
\ben
\chi (u,\brho)= r^{d/2}\phi(r,t)
\label{2-33}
\een
The full equation of motion becomes
\ben
-2\partial_u\partial_\brho \chi = \cP \chi + \bG (\brho) \chi^3
\label{2-34}
\een
where
\bea
\cP & \equiv & -\partial_{\brho}^2 + \bV_0 (\brho) \nn \\
\bV_0(\brho) & \equiv & r^2 f_b(r) \left[ \frac{d}{2}  r \partial_r f_b(r) + \frac{d(d+2)}{4} f_b(r) + m^2 \right] \nn \\
\bG(\brho) &  \equiv &  \frac{f_b(r)}{r^{d-2}} \nn \\
\label{2-35}
\eea
In the following we will use this form of the equations of motion to
examine the dynamics.

The discussion of multicritical points is exactly the same as that in
the $AdS$ soliton background in the previous section and will not be
repeated here.

\section{Slow Quench with a time dependent $\kappa$ in $AdS_{d+2}$ soliton background}

We now study the response of the system in the $AdS_{d+2}$ soliton background to a time dependent $\kappa$
which starts off {\em slowly} at early times in the ordered phase
$\kappa_i < \kappa_c$, crosses $\kappa_c$ and asymptotes at late times
for some other value $\kappa_f > \kappa_c$. The details of the
protocol are not important - however the manner in which the coupling
crosses the critical value is relevant. We consider a quench which is
linear near $\kappa \sim \kappa_c$, though all the considerations can
be trivially extended to nonlinear quenches. For concreteness we
consider the protocol
\ben \kappa (t) = \kappa_c + a \tanh(vt)
\label{3-1}
\een
with $v \ll 1$. Note that we are using units $r_0 = 1$  so what we
really mean is that $v \ll r_0$.

\subsection{Breakdown of Adiabaticity}

At early times, the response of the system is adiabatic. The
solution to the equation of motion (\ref{2-16a}) can be then obtained
in an adiabatic expansion \ben \psi(\rho,t;\kappa) =
\psi_0(\rho;\kappa(t))+ \epsilon \psi_1(\rho,t;\kappa) + \epsilon^2
\psi_2 (\rho,t;\kappa) + \cdots ,\label{3-2} \een where the static
solution is denoted by $\phi_0(r;\kappa)$ and $\epsilon$ is
the adiabaticity parameter. In the left hand side of equation
(\ref{2-16a}) we now need to replace $\partial_t \rightarrow \epsilon
\partial_t$ and equate terms with the same power of $\epsilon$. The
$n$-th order contribution to the solution, $\phi_n$ satisfies a {\em
  linear, inhomogeneous} equation with the source being determined by
the $\phi_m$ with $m < n$.

For the $\phi^4$ theory the two lowest order corrections satisfy
\bea \left[
  \cD + 3G(\rho) \psi_0^2 \right] \psi_1 & = & 0 \nn \\ \left[ \cD +
  3G(\rho) \psi_0^2 \right] \psi_2 & = & -\partial_t^2 \psi_0 -
3G(\rho) \psi_1^2 \psi_0
\label{3-3}
\eea
Note that $\psi_n$ for $n > 0$ satisfy vanishing boundary conditions
at infinity and regularity conditions in the interior. Since $\psi_1$
satisfies a homogeneous equation there is no nontrivial solution. The
lowest order correction to the adiabatic solution is therefore
$\psi_2$
\ben \psi_2 (\rho,t;\kappa) = - \int_0^{\rho_\star}
d\rho^\prime \cG (\rho, \rho^\prime) \partial_t^2 \psi_0
(\rho^\prime,\kappa(t))
\label{3-4}
\een
where $\cG (\rho, \rho^\prime)$ is the Green's function of the
operator $\left[ \cD + 3G(\rho) \psi_0^2 \right]$.

Exactly at $\kappa = \kappa_c$ the operator $\cD$ has a zero mode. At
this point the Green's function diverges and adiabaticity fails. As
found in the previous section, the smallest eigenvalue for a $\kappa$
close to $\kappa_c$ is proportional to $(\kappa -
\kappa_c)$. Furthermore we also found that $\psi_0 \sim (\kappa_c -
\kappa)^{1/2}$. Thus the lowest eigenvalue of the entire operator
$\left[ \cD + 3G(\rho) \psi_0^2 \right]$ is proportional to $(\kappa_c
- \kappa)$. This gives an estimate of $\psi_2$ as we approach the
critical point,
\ben
\psi_2 \sim \frac{1}{\kappa_c - \kappa
  (t)}\partial_t^2 \sqrt{\kappa_c - \kappa(t)} =
\frac{1}{2(\kappa_c-\kappa)^{3/2}}\left[ \partial_t^2 \kappa(t) +
  \frac{(\partial_t \kappa)^2}{\kappa_c - \kappa (t)}\right]
\label{3-5}
\een
The adiabatic expansion breaks down once $\psi_2 \sim \psi_0$ which leads to the condition
\ben \left[ \partial_t^2 \kappa(t) + \frac{(\partial_t
    \kappa)^2}{\kappa_c - \kappa (t)}\right] \sim (\kappa_c -
\kappa)^2 \label{3-6} \een For the protocol like (\ref{3-1}), or any
other protocol which is linear in time as it crosses $\kappa_c$ this
leads to the estimate for the time when adiabaticity fails, $t_{ad}$
\ben t_{ad} \sim v^{-1/3} \label{3-7} \een At this time the value of
the order parameter is then \ben \langle\cO\rangle \sim
(vt_{ad})^{1/2} \sim v^{1/3} \label{3-8} \een
This analysis can be easily repeated for multicritical points - this will be discussed in detail in a separate section.

\subsection{Dynamics in the critical region}

Once adiabaticity is broken there is no power series expansion in
$v$. We will now show that there is nevertheless an expansion for
small $v$, but in fractional powers of $v$. To see this let us
rescale
\ben \psi (\rho,t) = v^{1/3} \varphi (\rho,t)~~~~~~~~t =
v^{-1/3}\eta \label{3-9}
\een
The equation of motion (\ref{2-16})
now becomes
 \ben \cD \varphi = v^{2/3} \left(- \partial_\eta^2
\varphi - G(\rho) \varphi^3 \right) \label{3-10}
\een
Now decompose
the field in terms of eigenfunctions of $\cD$
\bea
\varphi (\rho,\eta) & = & \sum_n \chi_n(\rho) \xi_n(\eta) \nn \\
\cD \chi_n &  = & \lambda_n (\kappa) \chi_n \label{3-11} \eea The
eigenvalues of course depend on the boundary conditions. We  have
expressed this explicitly by denoting them by $\lambda_n(\kappa)$.
The equation (\ref{3-10}) becomes \ben \lambda_n (\kappa) \xi_n
(\eta) =v^{2/3} \left( - \partial_\eta^2 \xi_n - C^n_{m_1,m_2,m_3}
\xi_{m_1}  \xi_{m_2}  \xi_{m_3} \right) \label{3-12} \een where \ben
C^n_{m_1,m_2,m_3} \equiv \int_0^{\rho_\star} d\rho G(\rho)
\varphi_n^\star (\rho) \varphi_{m_1} (\rho)  \varphi_{m_2} (\rho)
\varphi_{m_3} (\rho) \label{3-13} \een In the previous section we
showed explicitly that the lowest eigenvalue of $\cD$ is of order
$(\kappa_c - \kappa)$. In fact generically for theories with
$\nu=1/2$ \ben \lambda_n(\kappa) = \lambda_n(\kappa_c) - c_n
(\kappa_c - \kappa) + O[(\kappa_c - \kappa)^2] ~~~~~~~c_n > 0
\label{3-13a} \een Using the fact that \ben \kappa_c - \kappa(t)
\sim-a (vt) = -av^{2/3} \eta \een in the critical region, the
equation ({3-12}) becomes \ben
 \lambda_n
(\kappa_c) \xi_n (\eta) =v^{2/3} \left( - \partial_\eta^2 \xi_n -a c_n \eta \xi_n - C^n_{m_1,m_2,m_3} \xi_{m_1}  \xi_{m_2}  \xi_{m_3} \right)
\label{3-14} \een The boundary condition gives rise to a time
dependent mass term in the equation for the mode functions. Recall
that $\lambda_0(\kappa_c) = 0$. The dominance of this zero mode for
small $v$ is manifest in this equation. All the other modes are at
least $O(v^{2/3})$. The zero mode satisfies an effective
Landau-Ginsburg dynamics, \ben
\partial_\eta^2 \xi_0 + c_0 \eta +  C^0_{000} \xi_0^3 = 0
\label{3-15} \een The order parameter, which is given in terms of
the asymptotic behavior of the field, also satisfies this equation
to lowest order. Reverting to the original variables we see that the
order parameter as a function of time has the scaling behavior \ben
\langle\cO\rangle(t;v) = v^{1/3} \langle\cO\rangle (tv^{1/3};1)
\label{3-16} \een
The dynamics is relativistic
and, as will be discussed in a later section,
consistent with $z=1$ Kibble Zurek scaling.

Once again the scaling solution for multicritical points follow
along similar lines, as discussed below.

\section{Slow quench with a time dependent $\kappa$: $AdS_{d+2}$ black brane}

The analysis for the response to a slow quench in a black brane
background is quite similar to above, but the results are rather
different.  We will not detail the analysis, but give the essential
equations, highlighting the results. The key difference arises from
the presence of a horizon in this geometry. We need to impose ingoing boundary conditions at the horizon. Equivalently, in the ingoing Eddington-Finkelstein
coordinates we are using we need to impose a regularity at the horizon
$r=1$ (in $\br_0=1$ units) \cite{hubeny}.

The time coordinate is now $u$, so that the protocol is \ben
\kappa(u) = \kappa_c + a \tanh (vu) \een Note that on the boundary
$u$ becomes the same as the usual time $t$ and in fact for any $r$
we have $\partial_t |_r = \partial_u |r$, so that on the boundary
this represents a time dependence identical to (\ref{3-1}).

\subsection{Breakdown of Adiabaticity}

Let us first discuss usual critical points ($\phi^4$ potential).
Since the equation of motion (\ref{2-34}) is first order in $u$
derivatives the first order correction to the adiabatic result is
non-vanishing. In the expansion \ben \chi(\brho,u;\kappa) =
\chi_0(\brho,\kappa(u))+ \epsilon \chi_1 (\brho,u) + \cdots
\label{4-1} \een the first order correction $\chi_1$ satisfies \ben
\left[ \cP + 3\bG(\brho) \chi_0^2 \right] \chi_1=-2\partial_u
\partial_\brho \chi_0 \label{4-2} \een An analysis identical to the
one between equations (\ref{3-3}) and (\ref{3-5}) then leads to \ben
\chi_1 \sim \frac{1}{\kappa_c - \kappa(u)}\partial_u \sqrt{\kappa_c
- \kappa(u)} \label{4-3} \een The condition $\chi_0 \sim \chi_1$
then leads to the adiabaticity breaking time \ben u_{ad} \sim
v^{-1/2} \label{4-4} \een while the expectation value of the
operator at this time is \ben \langle\cO \rangle \sim v^{1/4}
\label{4-5} \een The extension of these results to multicritical
points with the leading term in the
 potential being $\phi^{n+1}$ is straightforward, leading to
\ben u_{ad} \sim v^{-1/2}~~~~~~~\langle \cO \rangle \sim
v^{\frac{1}{2(n-1)}} =  v^{\beta/2} \label{4-6} \een We will
show below that these results are consistent with the
general Kibble-Zurek relations.

\subsection{Dynamics in The Critical Region}

For the $\phi^4$ theory we first rescale \ben \chi (\brho,u) = v^{1/4}
\tchi (\brho,\eta)~~~~~~~u = v^{-1/2}\eta
\label{4-7}
\een
so that the equation (\ref{2-34}) becomes
\ben
\cP \tchi = -v^{1/2} \left[2\partial_\brho\partial_\eta\tchi + \bG (\brho)
  \tchi^3 \right]
\label{4-8}
\een
Unlike the soliton, the spectrum of $\cP$ is now continuous. Therefore
the mode decomposition (\ref{3-11}) is replaced by
\ben
\tchi (\brho,\eta) = \int dk ~\tchi_k(\brho;\kappa) \bxi^k (\eta)
\label{4-9}
\een
where
\ben
\cP  \tchi_k(\brho;\kappa) = \blambda (k;\kappa) \tchi_k(\brho;\kappa)
\label{4-10}
\een so that instead of (\ref{3-12}) we get \ben \blambda(k;\kappa)
\bxi_k = -v^{1/2} \left[ \int dk^\prime \bB_{k k^\prime} \partial_\eta
  \bxi^{k^\prime} (\eta) + \int dk_1 dk_2
  dk_3~\bC^k_{k_1k_2k_3}\bxi^{k_1}\bxi^{k_2}\bxi^{k_3} \right]
\label{4-11}
\een
where
\bea
\bB_{k k^\prime} & = & \int d\brho~\tchi_k(\brho)\partial_\brho \tchi_{k^\prime} (\brho) \nn \\
\bC^k_{k_1k_2k_3} & = & \int d\brho~\bG(\brho)
\tchi_{k}(\brho)\tchi_{k_1}(\brho)\tchi_{k_2}(\brho)\tchi_{k_3}(\brho)
\label{4-12}
\eea

Since the operator $\cP$ is related to the operator $\cD$ in
(\ref{2-9}) by a similarity transformation (with the replacement
$\br_0 \rightarrow r_0$) the behavior of the eigenvalues
$\blambda(k;\kappa)$ near $\kappa = \kappa_c$ is the same as that of
$\lambda_n$ in (\ref{3-13a})
\ben \blambda(k;\kappa) =
\blambda(k;\kappa_c) - \bc(k)(\kappa_c - \kappa) + O((\kappa_c -
\kappa)^2)
\label{4-13}
\een
and using the time dependence of $\kappa (u)$ near $\kappa_c$ we get
\ben
\blambda(k;\kappa_c) \bxi_k = -v^{1/2} \left[ a \bc(k) \eta\bxi^k
  + \int dk^\prime \bB_{k k^\prime} \partial_\eta \bxi^{k^\prime}
  (\eta) + \int dk_1 dk_2
  dk_3~\bC^k_{k_1k_2k_3}\bxi^{k_1}\bxi^{k_2}\bxi^{k_3} \right]
\label{4-14}
\een
Recall that there is a zero mode at $\kappa = \kappa_c$ where the left
hand side of (\ref{4-14}) vanishes.  If the spectrum of
$\blambda(k;\kappa_c)$ were discrete it is clear from (\ref{4-14})
that the zero mode dominates the dynamics. This is what happens for
the $AdS$ soliton in the previous section. However the operator $\cP$
with $\kappa = \kappa_c$ has a continuous spectrum and one has to
careful. This analysis is, however, identical to that of
\cite{basu-das}.

The equation (\ref{4-14}) suggests a solution which is an expansion in
$v^{1/2}$ as follows
\ben
\bxi^k (\eta) = \delta(k) \txi^0 (\eta)+
v^{1/2}\txi^k(\eta) + O(v) \label{4-15}
\een
where again to lowest
order in small $v$
\bea 0 & = & \cB_{00}\partial_\eta \txi^0(\eta) +
a\bc_0\eta \txi^0(\eta) +\bC^0_{000} (\txi^0)^3 \nn \\ \txi^k & = &
-\frac{1}{\blambda(k;\kappa_c)} \left[ \cB_{k0}\partial_\eta
  \txi^0(\eta) + a\bc_k \eta \txi^0(\eta) +\bC^k_{000} (\txi^0)^3 \right]
\label{4-16}
\eea
Combining the two equations in (\ref{4-16}) one has
\ben \txi^k = -\frac{1}{\blambda(k;\kappa_c)} \left[
  (\cB_{k0}-\cB_{00}) \partial_\eta \txi^0(\eta) + a(\bc_k-\bc_0)\eta
  \txi^0(\eta) +(\bC^k_{000}-\bC^0_{000}) (\txi^0)^3 \right]
\label{4-17} \een We know that all the eigenvalues
$\blambda(k;\kappa_c)$ are positive except the one which is zero.
Since these positive eigenvalues form a continuum we can, without
loss of generality, write $\lambda(k;\kappa_c) = k^2$. This means
that our expansion (\ref{4-15}) is valid only if the quantities
$(\cB_{k0}-\cB_{00}), (\bc_k-\bc_0), (\bC^k_{000}-\bC^0_{000})$ go
to zero {\em at least as fast as $k^2$}. In a way quite similar to
\cite{basu-das} it turns out that this is indeed true - precisely
when $\kappa = \kappa_c$. This is shown in detail for a toy model
which is quite similar to our case in the appendix.

We therefore conclude that the dynamics in the critical region is
again dominated by the zero mode which now satisfies a
Landau-Ginsburg equation with a first order time derivative - the
first equation in (\ref{4-16}). This clearly yields a dissipative
time evolution. The dissipation is of course due to a finite
temperature and is caused by inflow into the horizon. Reverting to
the original variables and noting that on the boundary $u = t$, the
time of the field theory, we get a scaling solution 
\ben \label{4-80}
\langle\cO
\rangle(t;v) = v^{1/4} \langle\cO(tv^{1/2};1)\rangle 
\een This will
be shown to be consistent with Kibble-Zurek scaling with $z=2, \nu
=1/2$.

\section{Numerical Results}

\subsection{Soliton background}

After suitable changes of variables and field redefinitions for simplification, we solved the resulting equation of motion on a Chebyshev grid using pseudo-spectral derivative method. The $k$-th lattice point on a Chebyshev grid is defined in the following way,
\ben
\rho_k = \rho_\star \bigg( 1 - \cos \frac{k \pi}{N} \bigg)
\een
where, $N$ denotes the total number of points on the grid. At the center of the soliton we put a regularity condition on the field $\phi$.

We dealt with a specific case of the $AdS_{d+2}$ soliton, taking $d=3$ and $m^2 = -15/4$ on a grid with total number of points, $N =61$. Setting the mass parameter at the conformal value simplifies the numerics. The numerical calculation of the critical exponent involves following steps : 
\begin{itemize}
\item First we calculated $\kappa_c$ using the linear static equation and obtained $\kappa_c \approx -0.495129$. 
\item Next, we solved the non-linear static equation on the Chebyshev grid iteratively using a $\kappa = \kappa_c-a$ in the boundary condition. $a$ is an arbitrary constant chosen to be $a=0.1$.
\item The above field configuration was used to specify the initial conditions at some early time $t=-t_{max}$ in the full dynamic equation, which was solved using a time dependent $\kappa$-profile of the form $\kappa(t) = \kappa_c + a \tanh(vt)$. Near the phase transition point (i.e. $t=0$) $\kappa$ behaves linearly like $\kappa \approx \kappa_c+ a \, v t$.

\item  This was done for various values of $v$. Using small numerical values of $v$ we expect to find the system in a scaling regime. At time $t=0$ the value of the order parameter, $\langle \cO \rangle$ was numerically calculated from the solution and then the suitable plot [see Figure(\ref{OvsV})] was made to check the scaling.

\begin{figure}[htbp]
	 \centering \includegraphics[width=13cm]{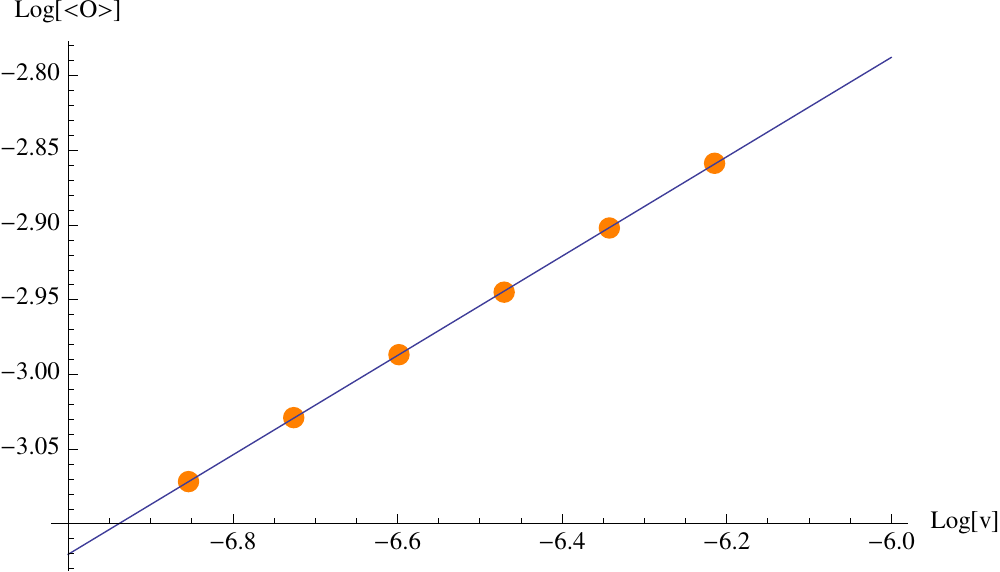}
	\caption{The scaling behavior of the order parameter $\cO$ as a function of $v$ in 
         in a  $\phi^4$ theory in $AdS$ soliton geometry. The fit gives, $\ln\langle \cO \rangle = -0.791971 + 0.332643 \ln v$. }
	\label{OvsV}
\end{figure}
\end{itemize}

The above fit clearly confirms our analytical expectation, viz., 
\ben
\langle \cO(0) \rangle \sim v^{1/3} 
\een
We also checked that changing $d \kappa$ and $N$ does not significantly change the exponent. To understand the full time dependence and the scaling of time (Eq. \ref{3-16}) one can plot the scaled response (Fig \ref{fig:Ovstsol}). 
\begin{figure}[htbp]
	 \centering \includegraphics[width=13cm]{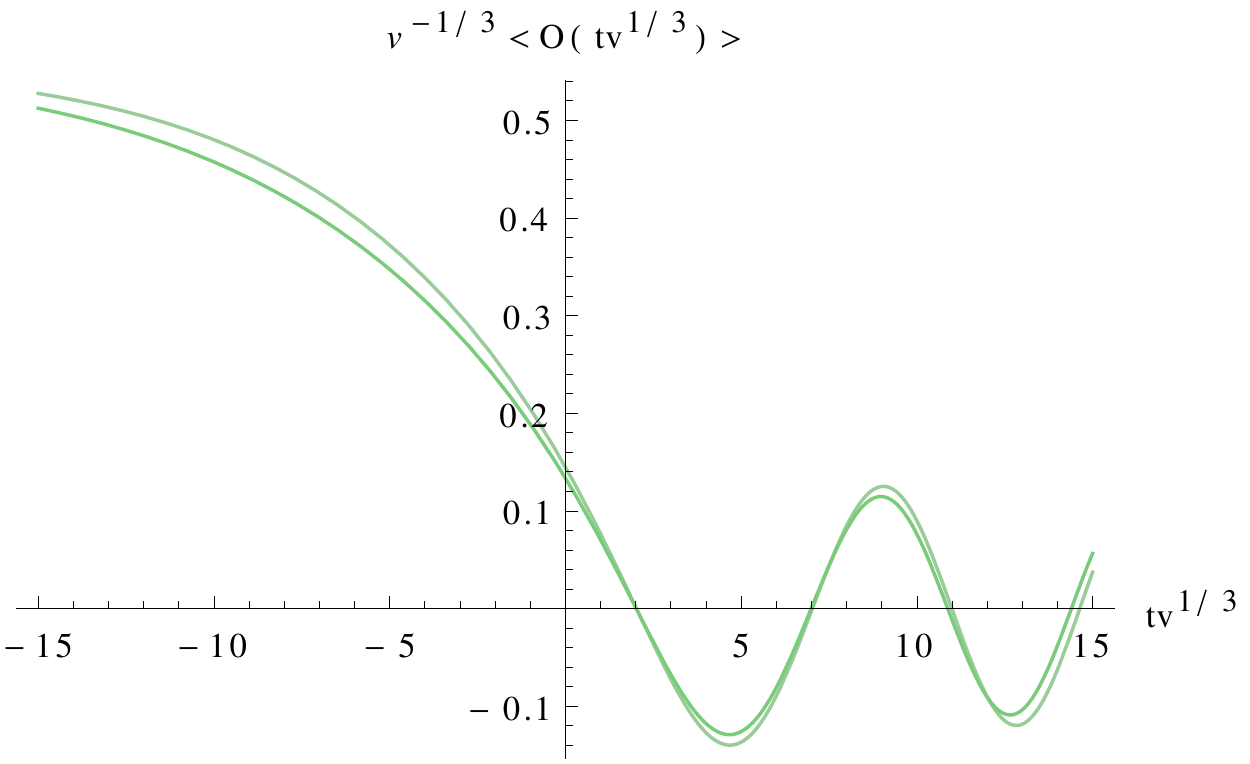}
	\caption{The scaling behavior of the order parameter $\cO$ as a function of $t$ in 
         in a  $\phi^4$ theory in $AdS$ soliton geometry.  Plots from the top are for $v=0.03,0.024$. These plots show scaling consistent with Eq. \ref{3-16}. }
	\label{fig:Ovstsol}
\end{figure}

\subsection{Black Brane background}

Here we solve the PDE's by slightly different method by calculating finite difference derivative on a lattice. We choose lattice size to be $npoints=500$. The resulting discretized equations are again solved by method of lines. Near the black hole horizon we impose an ingoing boundary condition. The main steps of the numerics, including the value of $\kappa_c$ and the time dependent profile $k(t)$, are identical to the soliton case : we do not repeat the details. The best fit here [see Figure(\ref{OvsVblk})] is given by,
$\ln\langle \cO \rangle =0.253967 \ln (v)-0.195079$ which conforms with our analytic result,
\ben
\langle \cO(0) \rangle \sim v^{1/4}.
\een
\begin{figure}[htbp]
	 \centering \includegraphics[width=13cm]{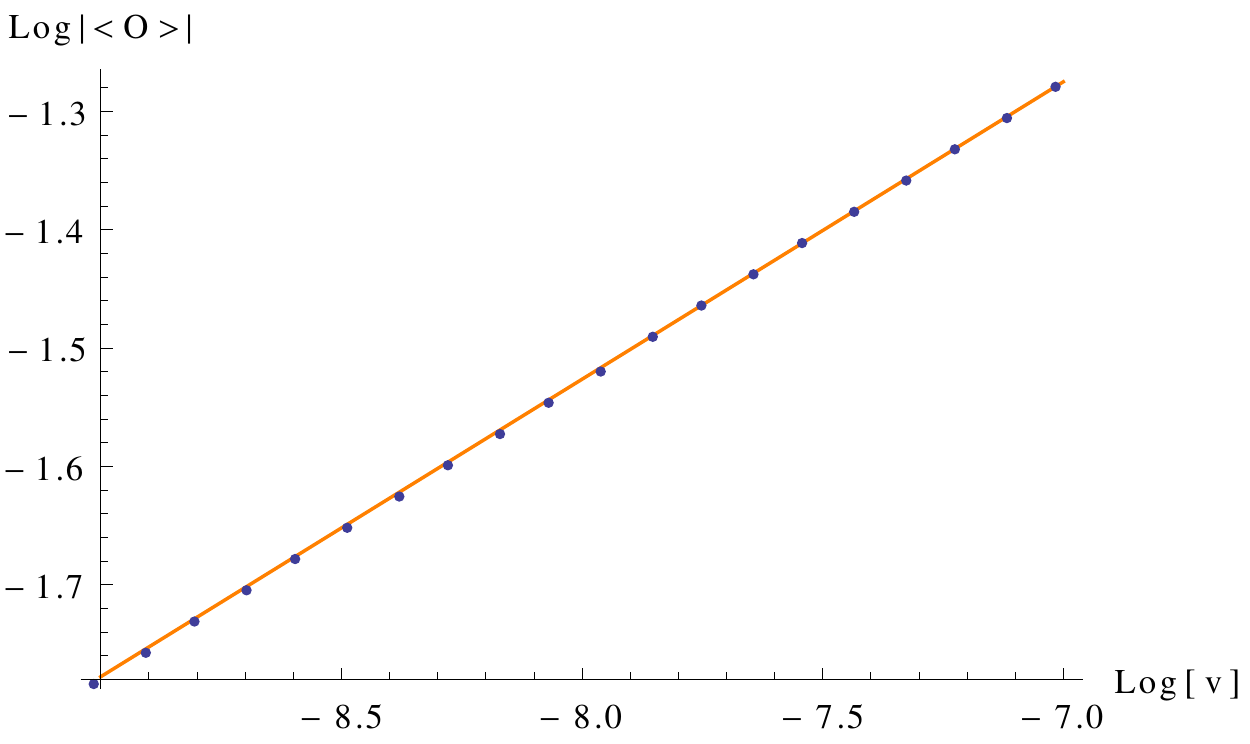}
	\caption{The scaling behavior of the order parameter $\cO$ as a function of $v$ for a
          $\phi^4$ theory in the $AdS$ blackhole background. The fit gives, $\ln\langle \cO \rangle =0.253967 \ln (v)-0.195079$. }
	\label{OvsVblk}
\end{figure}
Like the soliton case, we have also checked that the temporal scaling matches with Eq. \ref{4-80}.

In the probe approximation the late time behavior of the scalar field in black hole and soliton backgrounds are very different due to presence of the horizon in a black hole background. Any excess energy in bulk is gradually engulfed by the black hole and at very late time we have a almost static scalar profile. The late time decay of excitations of the scalar is determined by the quasi-normal modes. In a soliton background, the excess energy does not dissipate once the quenching is stopped and the scalar field shows temporal oscillation at late time. Our numerics confirm these assertions. 

\section{Arbitrary exponents and Kibble-Zurek Scaling}

In this section we discuss the connection of the holographic
{\em derivation} for scaling behavior in critical dynamics with the standard
{\em arguments} leading to Kibble Zurek scaling.

The standard argument for Kibble-Zurek scaling for a quantum critical
point proceeds as follows \cite{kibblezurek,sengupta}. In the vicinity
of such a transition the energy gap $\Delta$ depends on the control
parameter $\lambda$ (with the critical value of $\lambda$ chosen to
zero without loss of generality) as $\Delta \sim \lambda^{z\nu}$,
where $z$ is the dynamical critical exponent and the $\nu$ the
correlation length exponents. Consider quenching this control
parameter across this transition so that near the critical point
$\lambda \sim (vt)^{\alpha}$. Then the instantaneous value of the
energy gap in this region is given by $\Delta_{inst} \sim (vt)^{\alpha
  z \nu}$.
The criteria for the breakdown of adaibaticity during such a quench is
\cite{sengupta} $d \Delta/dt \sim \Delta^2$. Substituting the
expression for $\Delta$ and noting that the critical point is
reached at $t=0$, one finds that the time spent by the system in the
non-adiabatic regime is
\ben
T \sim v^{-\alpha z\nu/(\alpha z\nu+1)}
\label{5-1}
\een
Next one makes the important assumption that
the time evolution after the breakdown of adiabaticity is {\em
  diabatic}.
This means that one
can then argue that the order parameter is determined by the
instantaneous value at time $T$. Furthermore,
for slow quenches, the breakdown of adiabaticity occurs in the
critical region sufficiently close to the critical point so that one
can assume standard critical scaling holds.
Since in this region the order parameter $\cO$ varies
with the control parameter $\lambda$ as $\cO \sim \lambda^{\beta}$ we
get
 \ben
\langle\cO> \sim (v T)^{\alpha \beta} \sim v^{\alpha \beta/(\alpha
z\nu+1)} \label{5-2} \een

The adiabatic-diabatic assumption is rather drastic. 
In contrast, the holographic treatment of the present paper as well as that in
\cite{basu-das} and \cite{bddn} we {\em derived} a set of scaling
relations from the properties of the solutions to the bulk equations
of motion without any assumption about the nature of time evolution
after breakdown of adiabaticity.  The physics of the bulk is essential
in this derivation, which is not at all transparent in the boundary
field theory description. We will now show that the scaling relations
we obtained are, nevertheless, consistent with the standard Kibble
Zurek results described above.

There are several critical exponents involved in these
relations. First, the static exponent $\beta$ follows from the leading
nonlinearity of the bulk potential, as argued in section 2. If the
leading term in the potential is $\phi^{n+1}$ the value of $\beta$ is
given by equation (\ref{2-16d}), $\beta = 1/(n-1)$. To find the values
of $z$ and $\nu$ we need to look at the dispersion relation of small
fluctuations around the static solution. The linearized fluctuations
would satisfy an equation of the form
\ben
\partial_t^m \delta \psi = \left[ \cQ + nF(\rho) \psi_0^{n-1}
  \right]\delta \psi
\label{5-3}
\een
where $\psi_0$ is the static solution, and in the examples described
in this paper we have $m=2 , \cQ = \cD, F(\rho) = G(\rho)$ for the
soliton and
$m = 1, \cQ = \cP, F(\rho) = \bG (\brho)$ for
the black hole. The control parameter is $\lambda = (\kappa_c -
\kappa)$. The second term on the right hand side is therefore always
of the order $O(\lambda)$. The dependence of the first term on
$\lambda$ is determined by the nature of the background. Suppose the
smallest eigenvalue of $\cQ$ is $O(\lambda^{1/p})$. In both the soliton
and the black hole we had $p=1$ : here we have allowed for the
possibility of other backgrounds with arbitrary $p$. Then the energy
of excitations $E$ is given by
\bea
E^m & \sim & \lambda^{1/p}~~~~~~~~~p > 1 \nn \\
E^m & \sim & \lambda~~~~~~~~~~~~ p < 1
\label{5-5}
\eea
From the definition of the standard exponents we therefore have
\bea
z\nu & = & \frac{1}{pm}~~~~~~~~~p > 1 \nn \\
z\nu & = & \frac{1}{m}~~~~~~~~~~p < 1
\label{5-6}
\eea
Now consider the condition for breakdown of adiabaticity. This again
involves a solution of an equation of the form
\ben
\partial_t^m \psi_0 = \left[ \cQ + nF(\rho) \psi_0^{n-1}
  \right]\psi^\prime
\label{5-3a}
\een
where $\psi^\prime$ denotes the leading correction to the adiabatic
result. It is then clear that the condition $\psi^\prime \sim \psi_0$
leads to an adiabaticity breaking time $t_{ad}$
\bea
t_{ad} & \sim (v)^{-\frac{\alpha}{\alpha + pm}}~~~~~~p > 1 \nn \\
t_{ad} & \sim (v)^{-\frac{\alpha}{\alpha + m}}~~~~~~p < 1
\label{5-7} \eea With the value of $(z\nu)$ obtained in (\ref{5-6})
this reproduces the relation (\ref{5-1}). The instantaneous value of
$\psi$, and therefore the order parameter at this time is then
clearly seen to be \bea \langle\cO \rangle(t_{ad}) & \sim &
(v)^{\frac{m\alpha}{(n-1)({\alpha+m})}} = \cO_{ad}~~~~p >
1 \nn \\
\langle\cO \rangle(t_{ad}) & \sim &
(v)^{\frac{mp\alpha}{(n-1)({\alpha+mp})}} = \cO_{ad}~~~~p < 1
\label{5-8} \eea The value of $z$ requires knowledge of the way
space derivatives appear in the equation of motion. In the examples
we have discussed in this paper (as well as in \cite{basu-das,bddn})
the bulk equation of motion contains two space derivatives. Thus
with $m$ time derivatives we have $z=2/m$. It is then clear that
(\ref{5-8}) reproduces (\ref{5-2}) with $\beta = 1/(n-1)$ as derived
above.

Once the scalings of $t_{ad}$ and $\cO$ are known the rescaling of
fields and time required to expose the dynamics in the critical
region after breakdown of adiabaticity is clear - we need to perform
\ben t  \rightarrow \eta = \frac{t}{t_{ad}} ~~~~~~~~\psi \rightarrow
\chi = \frac{\psi}{\cO_{ad}} \label{5-9} \een The analysis of
sections (3.1) and (4.2) can be carried out in a straightforward
fashion leading to a scaling solution \ben \langle\cO\rangle(t,v) =
(v)^{\frac{mp\alpha}{(n-1)({\alpha+mp})}}
\langle\cO\rangle(t/t_{ad},1) \een which agrees with the
Kibble-Zurek solution obtained earlier.

In the above discussion we have indicated what should be the nature of the bulk theory which leads to nontrivial values of $z$ and $\nu$.
In a relativistic bulk theory, we always start with two time derivatives in the equations of motion. However the presence of a gauge field and/or a black hole horizon effectively leads to $m=1$. Values of $m \geq 3$ would be rather pathological in a bulk theory.

\section{Remarks}

As in \cite{basu-das} and \cite{bddn} we have demonstrated the
emergence of a scaling solution in the critical region in the probe
approximation for a quench which is more natural from the boundary
field theory point of view. The next obvious step is to study this issue with
gravitational backreaction, particularly for the soliton background. Pretty
much like global $AdS$ we expect that for a {\em slow} quench which does not
come close to a critical point, a black hole is not formed immediately \cite{bmin}.
A black hole may, however, form a late times \cite{bizon}. However near the critical point we expect that a breakdown of adiabatic evolution leads to a black hole formation at early times, and it would be interesting to look for critical behavior in this collapse.
This would involve coupled partial
differential equations - nevertheless we expect that the zero mode
will continue to play a key role and dominate critical dynamics. In
the full problem, however, it is important to consider potentials
which follow from a superpotential \cite{fhr,fhr2} so that the static solution is
stable. However once again we expect that near the critical point the
leading non-linearity determines the dynamics. These issues are being explored at the moment.

\section{Acknowledgements}
We would like to thank Robert Myers for a discussion. D.D. and
S.R.D. would like to thank International Center for Theoretical
Sciences of Tata Institute of Fundamental Research for hospitality
during the final stages of this work. The work of D.D. and S.R.D. was
supported in part by National Science Foundation grants
NSF-PHY/0970069 and NSF-PHY/1214341.

\section{Appendix : Validity of the small $v$ expansion}
\vspace{0.5cm}

To argue for the small $v$ expansion of $\bar{\xi}^k(\eta)$ (\ref{4-15})) we need to consider the eigenvalue
problem \ben [-\partial_\brho^2+ V_0(\brho)]\chi_k = k^2 \chi_k
\label{7egen}
\een
The above potential $V_0(\brho) \rightarrow
-e^{-\brho}$ as $\brho \rightarrow \infty$.

The basic features of the eigenfunctions can be understood from a
simpler problem in which we replace the potential by the following
potential which has the same qualitative features.  \ben
U(\brho) = \begin{cases} V_0  & \mbox{for } \brho = 0 \\
-U_0 & \mbox{for } 0 \leq
\brho \leq 1 \\ 
0 & \mbox{for } 1 \leq \brho \leq \infty
\end{cases}
\label{7squarewell}
\een
with $U_0, V_0 > 0$.
The eigenfunctions of the
Schrodinger operator with eigenvalue $k^2 > 0$ are \bea \psi_k(\brho) &
= & \frac{A(k)}{\sqrt{\pi}}\bigg( \sin (\sqrt{k^2+U_0}~\brho) + \kappa \cos (\sqrt{k^2+U_0}~\brho) \bigg)~~~~~~~~~~0
\leq \brho \leq 1 \nn \\ \psi_k(\brho) & = &
\frac{1}{\sqrt{\pi}}\sin(k\brho + \theta(\brho)) ~~~~~~~~~~1 \leq \brho
\leq \infty
\label{712-1}
\eea
Here $\kappa$ plays the role of the double trace deformation of our original problem, in the spirit that here too it dictates the modified boundary condition at $\brho = 0$. The constants $A(k)$ and $\theta(k)$ are determined by matching at $\brho = 1$,
\bea
A(k) &= & \frac{k}{\sqrt{k^2 (1+\kappa^2) +
   \bigg(  \cos(\sqrt{k^2+U_0})-\kappa \sin (\sqrt{k^2+U_0})\bigg)^2U_0}}\nn \\ \theta(k)&=&
\cos ^{-1}\left(\frac{ \bigg(  \cos(\sqrt{k^2+U_0})-\kappa \sin (\sqrt{k^2+U_0})\bigg)\sqrt{k^2+U_0}}{
 \sqrt{k^2 (1+\kappa^2) +
   \bigg(  \cos(\sqrt{k^2+U_0})-\kappa \sin (\sqrt{k^2+U_0})\bigg)^2U_0}}\right)-k.
\label{712-2}
\eea
The solution for $k=0$ is
\bea
\psi_0(\brho) & = & \frac{B}{\sqrt{\pi}} \bigg( \sin (\sqrt{U_0}~\brho)+\kappa \cos (\sqrt{U_0}~\brho) \bigg) ~~~~~~~~~~0 \leq \brho \leq 1 \nn \\
\psi_0(\brho) & = & \frac{1}{\sqrt{\pi}} \left( a\brho + b \right) ~~~~~~~~~~~~~~1 \leq \brho \leq \infty
\label{712-3}
\eea
The matching conditions at $\brho = 1$ now yield
\bea
B\left( \sin (\sqrt{U_0}) +\kappa \cos (\sqrt{U_0})\right) & = & a+b \nn \\
B\sqrt{U_0} \left( \cos (\sqrt{U_0}) -\kappa \sin (\sqrt{U_0} \right) & = & a
\label{712-4}
\eea
For any $a \neq 0$ the solution blows up at $\brho = \infty$. Thus regular solutions require $a = 0$. However the second equation in (\ref{712-4}) then imply that
\ben
\sqrt{U_0} = \cot^{-1} \kappa
\label{712-5}
\een
These are the zero modes. In the context of our model this is the potential where we have a critical point.

The small $k$ behavior of $A(k)$ and $\theta (k)$ can be read off from the expressions (\ref{712-2}). For a generic $U_0$ these are
\bea
A(k) & \sim & \frac{k}{ \left( \cos (\sqrt{U_0}) -\kappa \sin (\sqrt{U_0} \right)\sqrt{U_0}}+ O(k^3) \nn \\
\theta (k) & \sim & k [ \frac{\kappa + \tan \sqrt{U_0}}{\sqrt{U_0}\left( 1- \kappa \tan \sqrt{U_0}\right)} - 1] +O(k^3)
\label{712-6}
\eea
whereas for critical potentials we have
\bea
A(k) & \sim & \frac{1}{\sqrt{1+\kappa^2}}\left( 1 - \frac{k^2}{8} + O(k^4) \right)\nn \\
\theta (k) & \sim & \frac{\pi}{2} - \frac{k}{2}
\label{712-7}
\eea
This has implications for the coefficients like $(\cB_{k0}-\cB_{00})$ and $(\bC_{k000}-\bC_{0000})$ of equation(\ref{4-17}). Consider the quantity, $\cB_{k0}$. We have,
\ben
\cB_{k0}  =  \int d\brho~\tchi_k(\brho)\partial_\brho \tchi_{0} (\brho) \nn \\
\een
If we replace the true eigenfunctions by those of our simplified problem, we get
\bea
\cB_{k0} &=& A(k) \int_0^1 d\brho~ \bigg( \sin (\sqrt{k^2+U_0}~\brho) + \kappa \cos (\sqrt{k^2+U_0}~\brho) \bigg) B \sqrt{U_0} \nn \\ &\times& \left( \cos(\sqrt{U_0}\brho) -\kappa \sin (\sqrt{U_0}\brho) \right)
\eea
Using (\ref{712-6}) and (\ref{712-7}) we therefore see that
\ben
\cB_{k0}-\cB_{00} \sim k~~~~~~~k \rightarrow 0
\een
for generic potentials, whereas
\ben
\cB_{k0}-\cB_{00} \sim k^2~~~~~~~k \rightarrow 0
\een
for critical potentials. The behavior for $(\bC_{k000}-\bC_{0000})$ is similar.

Going back to (\ref{4-17}) we therefore see that
the small $v$ expansion is generically {\em not valid} since the
corrections diverge at small $k$. However for the critical potential,
$\tilde\xi_k$ remain finite as $k \rightarrow 0$ and the expansion in
powers of $v^{1/2}$ makes sense.


\begin{thebibliography}{99}



\bibitem{sengupta}
For reviews and references see
S.~Mondal, D.~Sen and K.~Sengupta,
Quantum Quenching, Annealing and Computation, Lecture Notes in Physics, Volume 802, arXiv:0908.2922,
 ;
J. ~Dziarmaga,Advances in Physics, vol. 59, issue 6, pp. 1063-1189,
arXiv:0912.4034 ;
  A.~Polkovnikov, K.~Sengupta, A.~Silva and M.~Vengalattore,
  Rev.\ Mod.\ Phys.\  {\bf 83}, 863 (2011)
  [arXiv:1007.5331 [cond-mat.stat-mech]].

\bibitem{CC}
  P.~Calabrese and J.~L.~Cardy,
  J.\ Stat.\ Mech.\  {\bf 0504} (2005) P010
  [arXiv:cond-mat/0503393];
  P.~Calabrese and J.~L.~Cardy,
  Phys.\ Rev.\ Lett.\  {\bf 96} (2006) 136801
  [arXiv:cond-mat/0601225];
  P.~Calabrese and J.~Cardy,
  [arXiv:0704.1880 [cond-mat.stat-mech]];
S.~Sotiriadis and J.~Cardy,
 J.\ Stat.\ Mech.\  (2008) P11003,
 [arXiv:0808.0116 [cond-mat.stat-mech]];
 S.~Sotiriadis, P.~Calabrese and J.~Cardy,
 EPL {\bf 87 }(2009) 20002,  [arXiv:0903.0895 [cond-mat.stat-mech]];




\bibitem{Maldacena}
  J.~M.~Maldacena,
  Adv.\ Theor.\ Math.\ Phys.\  {\bf 2} (1998) 231
  [Int.\ J.\ Theor.\ Phys.\  {\bf 38} (1999) 1113]
  [arXiv:hep-th/9711200];
S.~S.~Gubser, I.~R.~Klebanov and A.~M.~Polyakov,
  Phys.\ Lett.\ B {\bf 428}, 105 (1998)
  [arXiv:hep-th/9802109];
E.~Witten,
  Adv.\ Theor.\ Math.\ Phys.\  {\bf 2}, 253 (1998)
  [arXiv:hep-th/9802150];
O.~Aharony, S.~S.~Gubser, J.~M.~Maldacena, H.~Ooguri and Y.~Oz,
  Phys.\ Rept.\  {\bf 323}, 183 (2000)
  [arXiv:hep-th/9905111].

\bibitem{thermalization1} R.~A.~Janik and R.~B.~Peschanski,
 Phys.\ Rev.\  D {\bf 74}, 046007 (2006)
   [arXiv:hep-th/0606149];
 R.~A.~Janik,
   Phys.\ Rev.\ Lett.\  {\bf 98}, 022302 (2007)
   [arXiv:hep-th/0610144];
 P.~M.~Chesler and L.~G.~Yaffe,
   Phys.\ Rev.\ Lett.\  {\bf 102}, 211601 (2009)
   [arXiv:0812.2053 [hep-th]];
 P.~M.~Chesler, L.~G.~Yaffe,
   Phys.\ Rev.\  {\bf D82}, 026006 (2010).
   [arXiv:0906.4426 [hep-th]];
   S.~Bhattacharyya and S.~Minwalla,
   JHEP {\bf 0909} (2009) 034
   [arXiv:0904.0464 [hep-th]].
  D.~Garfinkle, L.~A.~Pando Zayas,
  Phys.\ Rev.\  {\bf D84}, 066006 (2011).
  [arXiv:1106.2339 [hep-th]];
  A.~Buchel, L.~Lehner and S.~L.~Liebling,
  Phys.\ Rev.\ D {\bf 86}, 123011 (2012)
  [arXiv:1210.0890 [gr-qc]].


\bibitem{otherthermalization}J.~Abajo-Arrastia, J.~Aparicio, E.~Lopez,
  JHEP {\bf 1011}, 149 (2010).
  [arXiv:1006.4090 [hep-th]];H.~Ebrahim, M.~Headrick,
  [arXiv:1010.5443 [hep-th]];
V.~Balasubramanian, A.~Bernamonti, J.~de Boer, N.~Copland, B.~Craps,
E.~Keski-Vakkuri, B.~Muller, A.~Schafer {\it et al.},
  Phys.\ Rev.\ Lett.\  {\bf 106}, 191601 (2011).
  [arXiv:1012.4753 [hep-th]];  V.~Balasubramanian, A.~Bernamonti, J.~de Boer,
N.~Copland, B.~Craps, E.~Keski-Vakkuri, B.~Muller, A.~Schafer {\it et al.},
  Phys.\ Rev.\  {\bf D84}, 026010 (2011).
  [arXiv:1103.2683 [hep-th]];
D.~Galante and M.~Schvellinger,
  JHEP {\bf 1207}, 096 (2012)
  [arXiv:1205.1548 [hep-th]] ;
E.~Caceres and A.~Kundu,
  JHEP {\bf 1209}, 055 (2012)
  [arXiv:1205.2354 [hep-th]].


\bibitem{holoentanglement}
T.~Albash, C.~V.~Johnson,
Electromagnetic Quenches,''
  New J.\ Phys.\  {\bf 13}, 045017 (2011).
  [arXiv:1008.3027 [hep-th]];
T.~Takayanagi, T.~Ugajin,
Coarse-Graining,''
  JHEP {\bf 1011}, 054 (2010).
  [arXiv:1008.3439 [hep-th]];
C.~T.~Asplund, S.~G.~Avery,
  [arXiv:1108.2510 [hep-th]].

\bibitem{dnt} S.~R.~Das, T.~Nishioka, T.~Takayanagi,
  JHEP {\bf 1007}, 071 (2010).
  [arXiv:1005.3348 [hep-th]].

\bibitem{otherapparent}
A.~O'Bannon,
arXiv:0808.1115 [hep-th].
 K.~Hashimoto, N.~Iizuka, T.~Oka,
  [arXiv:1012.4463 [hep-th]];
 K.~-Y.~Kim, J.~P.~Shock, J.~Tarrio,
  JHEP {\bf 1106}, 017 (2011).
  [arXiv:1103.4581 [hep-th]];
 S.~Prem Kumar,
  Phys.\ Rev.\  {\bf D84}, 026003 (2011).
  [arXiv:1104.1405 [hep-th]].
 S.~Janiszewski, A.~Karch,
[arXiv:1106.4010 [hep-th]];
 C.~Hoyos, T.~Nishioka, A.~O'Bannon,
 [arXiv:1106.4030 [hep-th]].

\bibitem{holocosmo}
  S.~R.~Das, J.~Michelson, K.~Narayan and S.~P.~Trivedi,
  Phys.\ Rev.\ D {\bf 74}, 026002 (2006)
  [hep-th/0602107];
  A.~Awad, S.~R.~Das, S.~Nampuri, K.~Narayan and S.~P.~Trivedi,
  Phys.\ Rev.\ D {\bf 79}, 046004 (2009)
  [arXiv:0807.1517 [hep-th]].
  A.~Awad, S.~R.~Das, A.~Ghosh, J.~-H.~Oh and S.~P.~Trivedi,
  Phys.\ Rev.\ D {\bf 80}, 126011 (2009)
  [arXiv:0906.3275 [hep-th]].


\bibitem{basu-das}
  P.~Basu and S.~R.~Das,
  JHEP {\bf 1201}, 103 (2012)
  [arXiv:1109.3909 [hep-th]].

\bibitem{kibblezurek}T. W. B. Kibble, J. Phys.A 9, 1387 (1976); W. H. Zurek,
Nature 317, 505 (1985).

\bibitem{bddn}
  P.~Basu, D.~Das, S.~R.~Das and T.~Nishioka,
  JHEP {\bf 1303}, 146 (2013)
  [arXiv:1211.7076 [hep-th]].


\bibitem{murata}K.~Murata, S.~Kinoshita and N.~Tanahashi,
  JHEP {\bf 1007}, 050 (2010)
  [arXiv:1005.0633 [hep-th]];


\bibitem{bhaseen}
  M.~J.~Bhaseen, J.~P.~Gauntlett, B.~D.~Simons, J.~Sonner and T.~Wiseman,
  Phys.\ Rev.\ Lett.\  {\bf 110}, 015301 (2013)
  [arXiv:1207.4194 [hep-th]].

\bibitem{buchel}
  A.~Buchel, L.~Lehner and R.~C.~Myers,
  JHEP {\bf 1208}, 049 (2012)
  [arXiv:1206.6785 [hep-th]].
  A.~Buchel, L.~Lehner, R.~C.~Myers and A.~van Niekerk,
  JHEP {\bf 1305}, 067 (2013)
  [arXiv:1302.2924 [hep-th]].
A.~Buchel, R.~C.~Myers and A.~van Niekerk,
  arXiv:1307.4740 [hep-th].

\bibitem{sotiriadis}  S.~Sotiriadis and J.~Cardy,
  arXiv:1002.0167 [quant-ph].

\bibitem{dsengupta} S.~R.~Das and K.~Sengupta,
  JHEP {\bf 1209}, 072 (2012)
  [arXiv:1202.2458 [hep-th]].

\bibitem{gubsersondhi} A. Chandran, A. Erez, S. Gubser and S. Sondhi, Phys. Rev.
B 86, 064304 (2012); A. Chandran, A. Nanduri, S. Gubser and S. Sondhi,
arXiv:1304.2402 [cond-mat.stat-mech]

\bibitem{mandalmorita}G.~Mandal and T.~Morita,
  arXiv:1302.0859 [hep-th].


\bibitem{liu1} N.~Iqbal, H.~Liu, M.~Mezei, Q.~Si,
  Phys.\ Rev.\  {\bf D82}, 045002 (2010).
  [arXiv:1003.0010 [hep-th]].

\bibitem{nishioka}
  T.~Nishioka, S.~Ryu and T.~Takayanagi,
  JHEP {\bf 1003}, 131 (2010)
  [arXiv:0911.0962 [hep-th]],
  G.~T.~Horowitz and B.~Way,
  JHEP {\bf 1011}, 011 (2010)
  [arXiv:1007.3714 [hep-th]].

\bibitem{fhr}
  T.~Faulkner, G.~T.~Horowitz and M.~M.~Roberts,
  JHEP {\bf 1104}, 051 (2011)
  [arXiv:1008.1581 [hep-th]].
\bibitem{fhr2}
  T.~Faulkner, G.~T.~Horowitz and M.~M.~Roberts,
  Class.\ Quant.\ Grav.\  {\bf 27}, 205007 (2010)
  [arXiv:1006.2387 [hep-th]].

\bibitem{alternative}
I.~R.~Klebanov and E.~Witten,
  Nucl.\ Phys.\ B {\bf 556}, 89 (1999)
  [hep-th/9905104];


\bibitem{doubletrace}
  E.~Witten,
  hep-th/0112258;
M.~Berkooz, A.~Sever and A.~Shomer,
  JHEP {\bf 0205}, 034 (2002)
  [hep-th/0112264];
  A.~Sever and A.~Shomer,
  JHEP {\bf 0207}, 027 (2002)
  [hep-th/0203168].
  W.~Mueck,
  Phys.\ Lett.\ B {\bf 531}, 301 (2002)
  [hep-th/0201100];
  S.~S.~Gubser and I.~R.~Klebanov,
  Nucl.\ Phys.\ B {\bf 656}, 23 (2003)
  [hep-th/0212138];
  L.~Vecchi,
  Phys.\ Rev.\ D {\bf 82}, 045013 (2010)
  [arXiv:1004.2063 [hep-th]];
  L.~Vecchi,
  JHEP {\bf 1104}, 056 (2011)
  [arXiv:1005.4921 [hep-th]].

\bibitem{hhh} S.~A.~Hartnoll, C.~P.~Herzog, G.~T.~Horowitz,
  JHEP {\bf 0812}, 015 (2008).
  [arXiv:0810.1563 [hep-th]];
  S.~A.~Hartnoll, C.~P.~Herzog, G.~T.~Horowitz,
  Phys.\ Rev.\ Lett.\  {\bf 101}, 031601 (2008).
  [arXiv:0803.3295 [hep-th]].

\bibitem{Basu:2013vva} 
  P.~Basu and A.~Ghosh,
  arXiv:1304.6349 [hep-th].

\bibitem{hubeny} 
  G.~T.~Horowitz and V.~E.~Hubeny,
  Phys.\ Rev.\ D {\bf 62}, 024027 (2000)
  [hep-th/9909056];
S.~Bhattacharyya, V.~EHubeny, S.~Minwalla and M.~Rangamani,
  JHEP {\bf 0802}, 045 (2008)
  [arXiv:0712.2456 [hep-th]].

\bibitem{bmin} 
  S.~Bhattacharyya and S.~Minwalla,
  JHEP {\bf 0909}, 034 (2009)
  [arXiv:0904.0464 [hep-th]].

\bibitem{bizon} 
  P.~Bizon and A.~Rostworowski,
  Phys.\ Rev.\ Lett.\  {\bf 107}, 031102 (2011)
  [arXiv:1104.3702 [gr-qc]];
O.~J.~C.~Dias, G.~T.~Horowitz, D.~Marolf and J.~E.~Santos,
  Class.\ Quant.\ Grav.\  {\bf 29}, 235019 (2012)
  [arXiv:1208.5772 [gr-qc]];
    D.~Garfinkle, L.~A.~Pando Zayas and D.~Reichmann,
  JHEP {\bf 1202}, 119 (2012)
  [arXiv:1110.5823 [hep-th]].

\bibitem{Kiritsis:2012ma} 
  E.~Kiritsis and V.~Niarchos,
  JHEP {\bf 1208}, 164 (2012)
  [arXiv:1205.6205 [hep-th]].

\end{thebibliography}
\end{document}